\documentclass[10pt, letterpaper]{article}

\RequirePackage[left=1in,right=1in,top=1in,bottom=1in]{geometry}

\usepackage{modular}
\usepackage{graphicx}
\usepackage{epstopdf}
\usepackage{color}
\usepackage{siunitx}
\usepackage{multirow}
\usepackage{algorithm}  
\usepackage{algorithmic}  
\usepackage{amsmath}
\usepackage{mathrsfs}
\usepackage{amsfonts}
\usepackage{hyphenat}
\usepackage{scrextend}
\usepackage{hyperref}
\usepackage{authblk}
\usepackage{cite}
\usepackage{amsthm}
\usepackage[utf8]{inputenc}

\title{Shannon-Limit Approached Information Reconciliation for Quantum Key Distribution}

\author[1$\dagger$]{Bang-Ying Tang}
\author[2$\dagger$$\ast$]{Bo Liu}
\author[1$\ast$]{Wan-Rong Yu}
\author[1]{Chun-Qing Wu}

\affil[1]{College of Computer, National University of Defense Technology, Changsha, 410073, China}

\affil[2]{College of Advanced Interdisciplinary Studies, National University of Defense Technology, Changsha, 410073, China}

\footnotetext[1]{Corresponding authors: liubo08@nudt.edu.cn, wlyu@nudt.edu.cn}
\footnotetext[2]{these authors contributed equally to this work}

\begin{document}
\maketitle
\begin{abstract}
Information reconciliation (IR) corrects the errors in sifted keys and ensures the correctness of quantum key distribution (QKD) systems. Polar codes-based IR schemes can achieve high reconciliation efficiency, however, the incidental high frame error rate decreases the secure key rate of QKD systems. In this article, we propose a \textbf{S}hannon-\textbf{l}imit \textbf{a}pproached (SLA) IR scheme, which mainly contains two phases: the forward reconciliation phase and the acknowledgment reconciliation phase. In the forward reconciliation phase, the sifted key is divided into sub-blocks and performed with the improved block checked successive cancellation list (BC-SCL) decoder of polar codes. Afterwards, only the failure corrected sub-blocks perform the additional acknowledgment reconciliation phase, which decreases the frame error rate of the SLA IR scheme. The experimental results show that the overall failure probability of SLA IR scheme is decreased to $10^{-8}$ and the efficiency is improved to 1.091 with the IR block length of \SI{128}{Mb}. Furthermore, the efficiency of the proposed SLA IR scheme is 1.055, approached to Shannon-limit, when quantum bit error rate is $0.02$ and the input scale of \SI{1}{Gb}, which is hundred times larger than the state-of-art implemented polar codes-based IR schemes. 
\end{abstract}
\newpage
\section{Introduction}

Quantum key distribution (QKD), can generate information-theoretical secure keys between distant communication parties (Alice and Bob)~\cite{RN4, RN162, RN163}. Assume the sifted keys are $K_\mathrm{s}^A$ and $K_\mathrm{s}^B$ with length of $n$ in both sides (Alice and Bob) after the quantum physical communication phase, $K_\mathrm{s}^A\neq K_\mathrm{s}^B$ with the quantum bit error rate (QBER) $E_\mu$, which introduced by imperfect implementations of QKD systems and potential attacks. Information reconciliation (IR), a critical procedure of the post-processing phase in QKD systems, aims at reconciling $K_\mathrm{s}^A$ and $K_\mathrm{s}^B$ to an equally weak secure key $K_\mathrm{IR}$, by exchanging the minimized extra syndrome information~\cite{RN4, RN145, RN148}. IR ensures the correctness of QKD systems and is the precondition to generate the final secure keys.

Initially, IR procedure is implemented performing interactive methods known as BBBSS~\cite{RN150,RN148} and Cascade~\cite{RN151}. Though high efficiency achieved by several improvements of Cascade algorithms, multiple rounds of communication are still required, resulting in significant heavy latency and authentication cost of QKD systems. Nowadays, IR is performed with forward error correction (FEC) codes, such as low-density parity-check (LDPC) codes~\cite{RN159,RN381} and polar codes~\cite{RN129, RN99, RN98}, where only one message contains a syndrome is exchanged between Alice and Bob, called as one-way IR scheme. Recently, most IR research focuses on performing with polar codes, for the advantage of the low computational complexity $O(n\log n)$ and high efficiency with potential to reach the Shannon limit, when the block size of a sifted key becomes as large as possible~\cite{RN122,RN125,RN124}. Though several improvements of polar decoders achieves higher IR efficiency with certain input scale ($\sim10^6$ bits), the correctness of QKD systems $\varepsilon$ is increased to the level of $10^{-3}$~\cite{RN129, RN98, RN96, RN99}. The state-of-art efficiency of polar codes-based IR scheme, reaches to 1.176 with the input block size of \SI{1}{Mb} when $E_\mu = 0.02$, while $\varepsilon$  still stays to $0.001$~\cite{RN99}. Actually, $\varepsilon$ should be decreased as low as possible (usually $<10^{-6}$), when performing polar codes into the IR procedure of QKD systems. 

Therefore, in this article, we propose a \textbf{S}hannon-\textbf{l}imit \textbf{a}pproached (SLA) IR scheme performing improved polar codes, which mainly composes of a forward reconciliation phase and an acknowledgment reconciliation phase. In the forward reconciliation procedure, a novel block checked successive cancellation list (BC-SCL) decoder was proposed to reduce the $\varepsilon$-correctness and error sub-blocks by remaining the successfully decoded sub-blocks with cyclic redundancy check (CRC) values in advance. Meanwhile, existed errors in sub-blocks after the forward reconciliation procedure, can be found by calculating the CRC values. For failure corrected sub-blocks, an additional acknowledgment reconciliation procedure is performed to decrease the $\varepsilon$ to the desired level. Finally, the corrected key $K_\mathrm{IR}$ is achieved. The experimental results show that our SLA IR scheme achieves correctness $\varepsilon$ to $10^{-8}$ and the reconciliation efficiency is better than 1.091 while the input block size is \SI{128}{Mb}. In principle, the efficiency and the SLA IR scheme can close to the Shannon-limit as the block length increases as large as possible. We achieved an efficiency of 1.055 with the $E_\mu=0.02$, when the input scale of SLA IR is increased to~\SI{1}{Gb}. Meanwhile, our SLA IR scheme with large-scale block size will benefit a lot in performing the rigorous statistical fluctuation analysis to remove the finite-size key effects~\cite{RN187,RN75} on the final secure key. Thus, SLA IR scheme can be efficiently implemented in practical QKD systems.

\section{Related Work}

\subsection{Information Reconciliation}
Information reconciliation (IR), as the critical post-processing procedure of QKD systems, corrects the errors in the sifted keys introduced by the implementation imperfectness and various attacks~\cite{RN4, RN115, RN146}, so as to ensure the correctness of QKD systems~\cite{RN378}. Assume the sifted key is $K_\mathrm{s}^A$ ($K_\mathrm{s}^B$) with length of $n$ on Alice's (Bob's) side, the quantum bit error rate (QBER) is $E_\mu$, the error corrected key is  $K_{\mathrm{IR}}^A$ and $K_{\mathrm{IR}}^B$, then the $\varepsilon-$correctness is equivalent to the requirement that the outputs of IR procedure, $K_{\mathrm{IR}}^A$ and $K_{\mathrm{IR}}^B$, differ only with small probability~\cite{RN378},

\begin{equation}
\mathrm{Pr}\left[K_\mathrm{IR}^A \neq K_\mathrm{IR}^B\right]\leq \varepsilon.
\end{equation}

Assume the key information learned by eavesdroppers is $S$, then the reconciliation efficiency is defined as

\begin{equation}
f\left(E_\mu\right) = \frac{1-\min\left\{H_2\left(K_{\mathrm{IR}}^A|S\right),H_2\left(K_{\mathrm{IR}}^B|S\right)\right\}}{H_2(E_\mu)},
\end{equation} 

where $H_2(x)$ is the binary Shannon entropy, calculated by

\begin{equation}
H_2\left(x\right) = -x\log_2\left(x\right) -\left(1-x\right)\log_2\left(1-x\right).
\end{equation} 

The average yield of IR scheme is given by

\begin{equation}
\label{eq:f}
\gamma=\left(1-\varepsilon\right)\min\left\{H_2\left(K_{\mathrm{IR}}^A|S\right),H_2\left(K_{\mathrm{IR}}^B|S\right)\right\}=(1-\varepsilon)\left[1-f\left(E_\mu\right)H_2(E_\mu)\right].
\end{equation}

\subsection{Polar codes-based IR schemes}

Given any binary-input discrete memoryless channel (B-DMC), E. Arikan first proposed a Shannon limit approached information reconciliation scheme with complexity $O\left(N\log N\right)$, named as polar codes in 2009~\cite{RN122,RN125}. In 2014, P. Jouguet and S. Kunz-Jacques performed the polar codes in the IR procedure in QKD systems, furthermore, they showed that polar codes have an equivalent efficiency below 1.12 for given upper bound $\varepsilon=0.1$ and block length starting from \SI{64}{Kb} to \SI{16}{Mb}~\cite{RN129}. Afterwards, A. Nakassis and A. Mink described flexible polar codes-based IR approaches for QKD systems and showed the potential to approach to the Shannon limit with a more efficient decoder when the location and values of the frozen bits were known at the design time~\cite{RN98}. S. Yan~\emph{et al.} improved the polar codes-based IR scheme with successive cancellation list (SCL) decoding and optimized coding structures, which decreased the $\varepsilon$ to the level of $10^{-3}$ and the equivalent efficiency reached to 1.176~\cite{RN99}. The detailed performance of above IR schemes is described in Table.\ref{otherwork}. 

\begin{table}[!htbp]
	\newcommand{\tabincell}[2]{\begin{tabular}{@{}#1@{}}#2\end{tabular}}%
	\centering
	\caption{The performance of polar codes-based IR schemes}\label{otherwork}%
	\begin{tabular}{|c|c|c|c|c|c|}
		\hline
		Author&QBER&$n$&  $f$ & $\varepsilon$ & $\gamma$\\
		\hline
		\multirow{3}*{P. Jouguet and S. Kunz-Jacques~\cite{RN129}}& \multirow{3}*{0.02}& \SI{64}{Kb} & 1.395& 0.090&0.731\\
		\cline{3-6}
		& & \SI{1}{Mb} &1.225& 0.110&0.736	\\
		\cline{3-6}
		& & \SI{16}{Mb}  & 1.121& 0.080& 0.774\\
		\hline
	\multirow{6}*{A. Nakassis and A. Mink~\cite{RN98}}& \multirow{2}*{0.02}& \SI{64}{Kb}  & 1.425& 0.073& 	0.740	\\
	\cline{3-6}
	& &  \SI{1}{Mb} & 1.243& 0.027&0.802\\
	\cline{2-6}
	& \multirow{2}*{0.04}& \SI{64}{Kb}  & 1.344& 0.015&	0.664\\
	\cline{3-6}
	& & \SI{1}{Mb}  & 1.188& 0.031&	0.690\\
	\cline{2-6}
	& \multirow{2}*{0.06}& \SI{64}{Kb} & 1.247& 0.068&	0.552\\
	\cline{3-6}
	& &  \SI{1}{Mb} & 1.144& 0.034&0.604\\
		\hline
		\multirow{2}*{S. Yan~\emph{et al.}~\cite{RN99}}& \multirow{2}*{0.02}& \SI{64}{Kb} & 1.261& 0.002&	0.820	\\
		\cline{3-6}
		& & \SI{1}{Mb} &  1.176& 0.001&0.833\\
		\hline
	\end{tabular}
\end{table}

\section{Shannon-limit approached IR scheme}

In principle, lower $\varepsilon$ and Shannon-limit $f$ of polar codes-based IR schemes can be approached with increased input block size and improved decoders~\cite{RN149,RN188,RN143,RN111,RN169,RN170,RN167,RN168}. Moreover, IR schemes with large-scale input block size benefit much in performing the rigorous statistical fluctuation analysis to remove the finite-size key effects on the final secure keys. However, $\varepsilon$ of state-of-art polar codes-based IR schemes still stays on the level of $10^{-3}$, which reduces the final secure key rates of QKD systems. 

\begin{figure}[htb]
	\centering
	\includegraphics[width=1\textwidth]{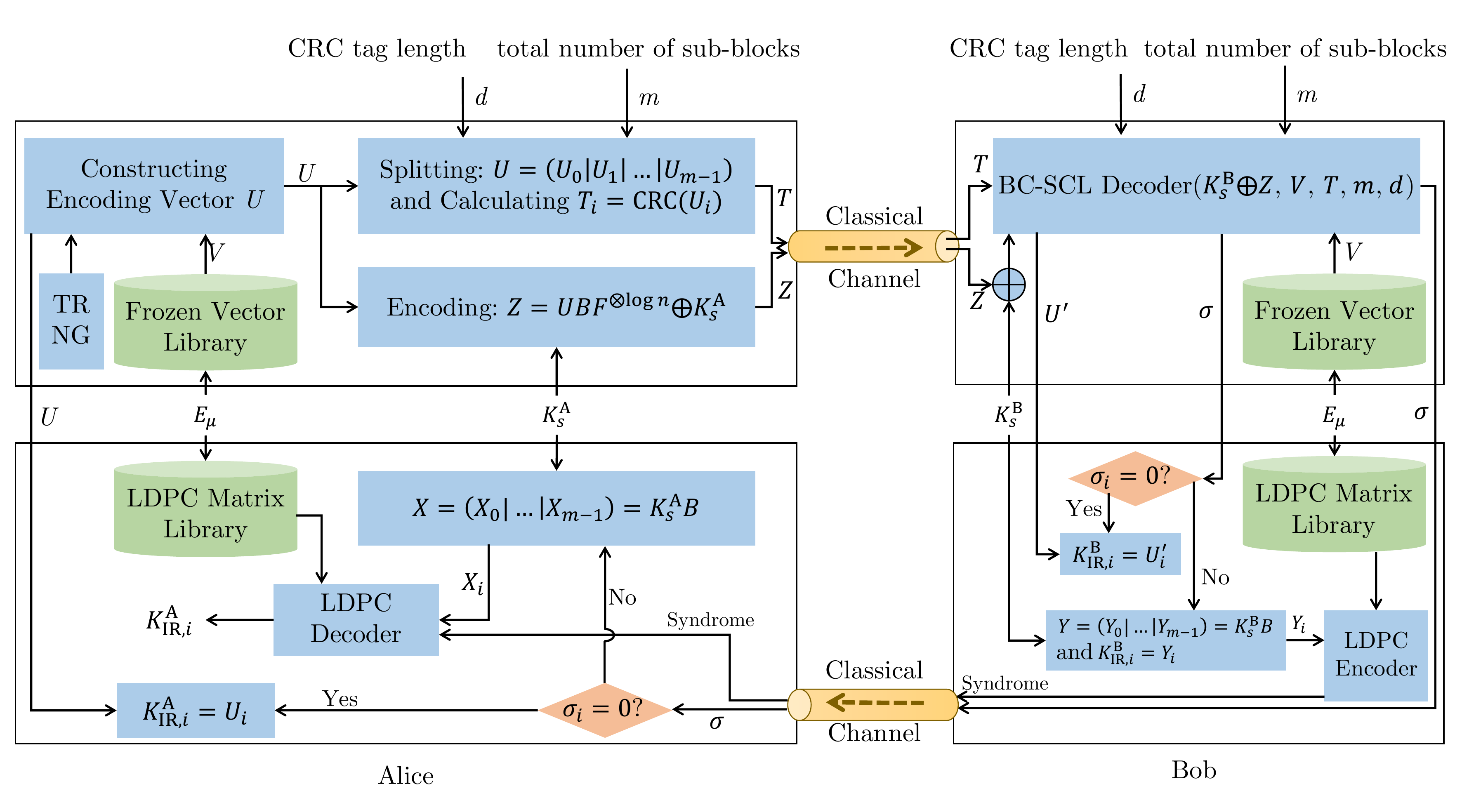}
	\caption{The schematic diagram of the proposed SLA IR scheme. TRNG: true random number generator, $E_\mu$: quantum bit error rate, CRC: cyclic redundancy check, $K_s^{\mathrm{A}}$ and $K_s^{\mathrm{B}}$: the sifted key of Alice and Bob, LDPC: low density parity check. The detailed description of this figure is shown in the main text.}
	\label{fig:mainpro}
\end{figure}

In this article, we propose an improved \textbf{S}hannon-\textbf{l}imit \textbf{a}pproached (SLA) IR scheme for QKD, and the schematic diagram is shown in Fig.~\ref{fig:mainpro}. The proposed SLA IR scheme mainly contains two phases: the forward reconciliation phase and the acknowledgment reconciliation phase. In the forward reconciliation phase, Alice constructs the encoding vector $U$ with true random numbers and the optimal frozen vector $V$, chosen from the frozen vector library with the quantum bit error rate $E_\mu$. Then, Alice calculates the syndrome $Z$ of $K_\mathrm{s}^A$ with the polar codes encoder, meanwhile, divides the vector $U$ to $m$ sub-blocks and calculates the cyclic redundancy check (CRC) value of each block. Afterwards, the syndrome $Z$ and combined CRC value vector $T$ are transmitted to Bob via classical channel. Bob performs the same operations to select the frozen vector $V$ and then performs the improved BC-SCL decoder (detail described in Section~\ref{subsec:BCC-SCL}) to get the corrected vector $U^\prime$ and the status vector $\sigma$, indicating which sub-lock is failure corrected. In the acknowledgment reconciliation phase, Alice and Bob performs a low density parity check (LDPC) error correction procedure to correct the error bits in the failure corrected sub-blocks. Afterwards, Alice and Bob obtain the uniform key $K_\mathrm{IR}$ respectively.

\subsection{Forward Reconciliation}

Before Alice and Bob start the SLA IR scheme, optimized multi-rate frozen vectors of polar codes and parity-check matrix of LDPC codes are shared between each other. 

First of all, Alice and Bob will calculate the required CRC length $d$ and choose the appropriate number of sub-blocks $m$ to achieve expected correctness $\varepsilon$. Given $E_\mu$, Alice selects the optimized frozen vector $V$ of polar codes, where frozen bits are set to ``0'' and the rest are set to ``-1''. Then, $k$ bits of true random numbers are used to replace the elements of $V$, whose value equals to ``-1'', marked as the vector $U$. Then, split $U$ to $m$ sub-blocks with length $n^\prime=n/m$. For each sub-block $U_i$, $i\in [0,m)$ and $i\in \mathbb{N}$, calculate the CRC tag value $T_i=\mathrm{CRC}\left(U_i\right)$ and combined to $T$, $T=(T_0|T_1|\dots|T_{m-1})$. 

Meanwhile, the vector $U$ is encoded to $Z$ by

\begin{equation}
Z=UG_n \oplus K_{\mathrm{s}}^A,
\end{equation}

where $G_n$ is the bit-reversal invariant matrix, defined as $G_n=BF^{\otimes \log n}$, $B$ is the permutation matrix for bit-reversal operation and  $F\overset{\Delta}{=} \begin{bmatrix} 1 & 0\\ 1&1 \end{bmatrix}$~\cite{RN122}. Afterwards, Alice sends $T$ and $Z$ to Bob via the classical channel.

At Bob's side, with Bob's sifted key $K_\mathrm{s}^B$, received $T$ and $Z$, we can get the decoded vector $U^\prime$ with failure probability $\varepsilon_\mathrm{f}$ by performing our improved novel block checked (BC) SCL decoder, detailed described in Section~\ref{subsec:BCC-SCL}. Additionally, a status vector $\sigma$ also given for indicating which sub-block $U^\prime_i$ is failure decoded. Each element of $\sigma$ is defined as

\begin{equation}
\sigma_i=\left\{\begin{matrix}
1 &  \mathrm{CRC}\left(U^\prime_i\right) \neq T_i \\ 
0 & else
\end{matrix}\right. .
\end{equation} 

Thus, in total $r$ sub-blocks are failure decoded, $r=\sum{\sigma_i}$. The position vector of these failure corrected sub-blocks is defined as $E \overset{\Delta}{=} \left\{i|\sigma_i=1\right\}$.

\subsection{Acknowledgment Reconciliation}

After the forward reconciliation phase, we have to perform the acknowledgment reconciliation phase to correct the remained errors in partial sub-blocks.

Here, Bob distinguishes two cases according to the value of $r$.

\textbf{\textit{Case I.}} If $r\neq 0$, Bob performs the permutation operation to $K_\mathrm{s}^B$, 

\begin{equation}
Y=K_\mathrm{s}^B B,
\end{equation}

then, divide $Y$ to $m$ sub-blocks with length $n^\prime=n/m$.  Then Bob calculates the syndrome $\mathscr{S} $ from $Y_E$ by performing the LDPC encoding scheme with chosen optimized parity-check matrix, where $Y_E\overset{\Delta}{=} \left\{ Y_{e_i}|e_i\in E \right\}$. Then, Bob sends $\sigma$ and $\mathscr{S}$ to Alice. Alice performs the bit-reversal operation to $K_\mathrm{s}^A$,

\begin{equation}
X=K_\mathrm{s}^A B,
\end{equation} 

then divide $X$ into $m$ sub-blocks with length $n^\prime=n/m$. Alice corrects the error bits in $X_E$ with $\mathscr{S}$, where $X_E\overset{\Delta}{=} \left\{ X_{e_i}|e_i\in E \right\}$. 

In the end of acknowledgment reconciliation phase, Bob gets the error corrected key $K_\mathrm{IR}^B$ with failure probability $\varepsilon$, whose $i$-th sub-block can be represented by

\begin{equation}
\label{equ:K_IR}
K_{\mathrm{IR},i}^B \overset{\Delta}{=} \left\{\begin{matrix}
U_i^\prime &  \sigma_i=0 \\ 
Y_i & \sigma_i\neq 0
\end{matrix}\right. .
\end{equation} 

Alice performs similar procedure shown in equation (\ref{equ:K_IR}) to the identical and weak secure key $K_\mathrm{IR}^A$.

\textbf{\textit{Case II.}} If $r=0$, Bob sets  $\mathscr{S} =\textrm \O$. Then, $\sigma$ and  $\mathscr{S}$ are transmitted to Alice. Afterwards, Alice and Bob gets $K_\mathrm{IR}^A=U$ and $K_\mathrm{IR}^B=U^\prime$ as the error corrected key, respectively.

\subsection{Block Checked SCL Decoder}
\label{subsec:BCC-SCL}

In the article, we improve the successive cancellation list (SCL) decoder to reduce the $\varepsilon$-correctness by performing cyclic redundancy check (CRC) to divided sub-blocks, called as block checked (BC) SCL decoder~\cite{RN149}. 

In the BC-SCL decoder, we assume the list size is $l$, $P$ is the list of decoded vectors, $\mathcal{P} \in P$ with length of $n$, $\mathcal{P}_{j}^{k}$ is a sub-vector of $\mathcal{P} \in P$, where $\mathcal{P}_{j}^{k}=[\mathcal{P}[j],\cdots,\mathcal{P}[k]]$, $0\leq j \leq k < n$ and the outcome of the decoder is $U$ which can be split into the sub-blocks $U_i$ of length $n'$, $0\leq i <m$. 

\newtheorem{myDef}{Definition}
\begin{myDef} $\mathcal{M}(\mathcal{P},i)$ is the path metric of the decoded vector $\mathcal{P}_{0}^{i}$, calculated as~\cite{RN392}
\begin{equation}
	\mathcal{M}\left(\mathcal{P},i\right)=-\ln \operatorname{Pr}\left(\mathcal{P}_{0}^{i}|K_\mathrm{s}^{B} \oplus Z\right),
\end{equation}
where $i\in\left[0,n\right)$.
\end{myDef}

\begin{myDef}
$\mathcal{M}_{\min}^{l}(P,i)$ is the $l$-th minimum path metric of $\mathcal{M}(P,i)$, where $\mathcal{M}(P,i) =  \left\{\mathcal{M}(\mathcal{P},i) | \mathcal{P} \in P\right\}$ and  $i\in\left[0,n\right)$.
\end{myDef}

\begin{myDef}
	\label{def_Fork}
$Fork(P,i)$ is defined as assume $P^\prime = P$, for $\forall \mathcal{P} \in P^\prime$, set $\mathcal{P}[i]=1$ and $P\leftarrow P \cup \mathcal{P}$, where $i\in\left[0,n\right)$~\cite{RN149}.
\end{myDef}

\begin{myDef}
	$Prune(P,i,l)$ is defined as the operation that when $\left| P \right|>l$, for $\forall \mathcal{P} \in P$ and $\mathcal{M}(\mathcal{P},i) > \mathcal{M}_{\min}^{l}(P,i)$, set $P\leftarrow P \setminus  {\mathcal{P}}$~\cite{RN392}.
\end{myDef}

The detailed description of the decoding procedure of BC-SCL decoder is shown in Algorithm.~\ref{algorithm_BCC}. 

\begin{algorithm}[htb]
	\caption{BC-SCL Decoder decoding procedure}
	\label{algorithm_BCC}
	\begin{algorithmic}[1]
		\REQUIRE{$l$, $n$, $m$, $T$, $V$, and $K_\mathrm{s}^{B} \oplus Z$} 
		\ENSURE{$U$, $\sigma$}
		\STATE{$n^\prime=n/m$, $U=0^n$}	
		\STATE{$P=\{0^n\}$}
		\FOR{$i=0$ to $m-1$}
		\FOR{$j=0$ to $n'-1$}
			\STATE{\textbf{if} $v_{in'+j}=-1$ \textbf{then} $\textit{Fork}(P,in'+j)$}
			\STATE{$\textit{Prune}\left (P,in'+j,l\right)$}
		\ENDFOR
		\STATE{\textbf{if} $\exists \mathcal{P}\in P, \mathrm{CRC}(\mathcal{P}^{(i+1)n'-1}_{in'})=T_i$ \textbf{then} $U_i=\mathcal{P}^{(i+1)n'-1}_{in'}$, $\sigma _{i}=0$}
		\STATE{\textbf{else} $\sigma _{i}=1$}
		\ENDFOR
		\STATE{\textbf{if} $\exists$ $\mathcal{P}\in P,$ $\forall i\in [0,m)$, $\mathrm{CRC}(\mathcal{P}_{in'}^{(i+1)n'-1})=T_i$ \textbf{then} $U=\mathcal{P}$}
	\end{algorithmic}
\end{algorithm}

\subsection{Performance of the SLA IR scheme}

Let $\mathcal{W}_i$ be the corresponding bit-channel of polar codes performed in our forward reconciliation phase of the SLA IR scheme, $P_e\left(\mathcal{W}_i\right)$ is the probability of error on the $i$th bit-channel, where $i=0,1,\cdots,n-1$. The union upper bound of correctness $\varepsilon_\mathrm{f}$ of forward reconciliation phase is estimated as~\cite{RN106}

\begin{equation}
\label{efficiency}
\varepsilon_\mathrm{f} \leq \sum_{i=0}^{n-1} -v_i P_e\left(\mathcal{W}_i\right).
\end{equation}

Then, we analyze the total correctness of the SLA IR scheme in two cases.

\textbf{\textit{Case I.}} $r\neq 0$. In this case, the total correctness $\varepsilon_\mathrm{I}$ can be calculated as
\begin{equation}
\varepsilon_\mathrm{I} \leq \varepsilon_\mathrm{f}\sum_{i=1}^{m}{\operatorname{Pr}\left(r=i\right)\left[1-\left(1-\frac{l}{2^d}\right)^{m-i}+\varepsilon_\mathrm{a}\right]}, 
\end{equation}
where $\varepsilon_a$ is the failure probability of the acknowledgment reconciliation phase and $1-\left(1-\frac{l}{2^d}\right)^{m-i}$ is the probability of error on $i$ sub-blocks which passed the CRC check in the forward reconciliation phase.

\textbf{\textit{Case II.}} $r = 0$. In this case, all outcome sub-blocks of the BC-SCL decoder will pass the CRC check in the forward reconciliation phase, and the total correctness $\varepsilon_\mathrm{II}$ can be calculated as

\begin{equation}
\varepsilon_\mathrm{II} \leq \varepsilon_\mathrm{f} \operatorname{Pr}\left(r=0\right)\left[ 1- \left(1-\frac{l}{2^d} \right) ^m \right].
\end{equation}

Thus, the total correctness $\varepsilon$ of SLA IR scheme can be calculated as
\begin{equation}
\label{correctness}
\begin{aligned}
	\varepsilon &\leq \varepsilon_\mathrm{f} \left \{ \operatorname{Pr}\left(r=0\right)\left[1-\left(1-\frac{l}{2^d}\right)^m\right]+\sum_{i=1}^{m}\operatorname{Pr} \left(r=i\right) \left[ 1-\left( 1-\frac{l}{2^d} \right) ^{m - i}+\varepsilon_\mathrm{a} \right] \right\}\\
	&< \varepsilon_\mathrm{f}\left\{ \operatorname{Pr}\left(r=0\right)\left[ 1-\left( 1-\frac{l}{2^d} \right)^m +\varepsilon _\mathrm{a} \right] +\sum_{i=1}^{m}\operatorname{Pr}\left(r=i\right) \left[ 1-\left( 1-\frac{l}{2^d}\right) ^{m}+\varepsilon_\mathrm{a} \right]\right\}\\
	&=\varepsilon_\mathrm{f} \left[1-\left(1-\frac{l}{2^{d}}\right)^{m}+\varepsilon_\mathrm{a} \right]
\end{aligned}.
\end{equation}

\begin{figure}[htb]
	\centering
	\includegraphics[width=0.8\textwidth]{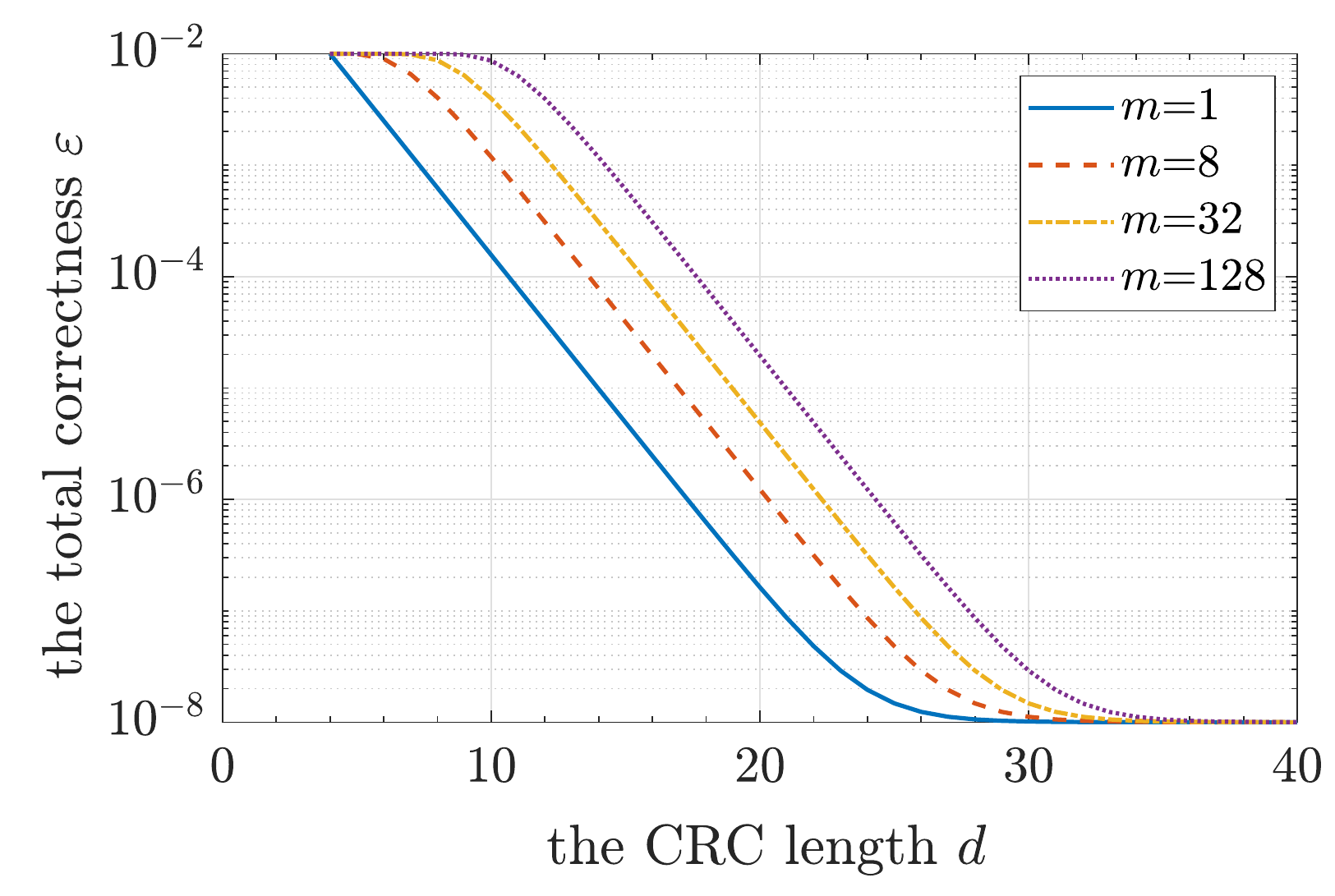}
	\caption{The correctness of the SLA IR scheme $\varepsilon$ versus $d$ with $l=16$, $m=1,8,32,128$ and $d\geq \log_2{l}$
		\label{fig:fer_sim}}
\end{figure}

With optimized construction of polar codes and LDPC codes~\cite{RN159,RN106}, we set $\varepsilon_\mathrm{a} \leq 10^{-6}$ and $\varepsilon_\mathrm{f} \leq 10^{-2}$. The analyzed results of $\varepsilon$ versus $d$ of the SLA IR scheme is shown in Fig.~\ref{fig:fer_sim}, according to equation  (\ref{correctness}), here $l=16$, $m=1,8,32,128$ and $d\geq \log_2{l}$. As shown in Fig.~\ref{fig:fer_sim}, the value of $\varepsilon$ becomes higher with larger $m$ and approaches to the lower bound of $10^{-8}$ when $d\geq 36$.

\begin{figure}[htb]
	\centering
	\includegraphics[width=0.8\textwidth]{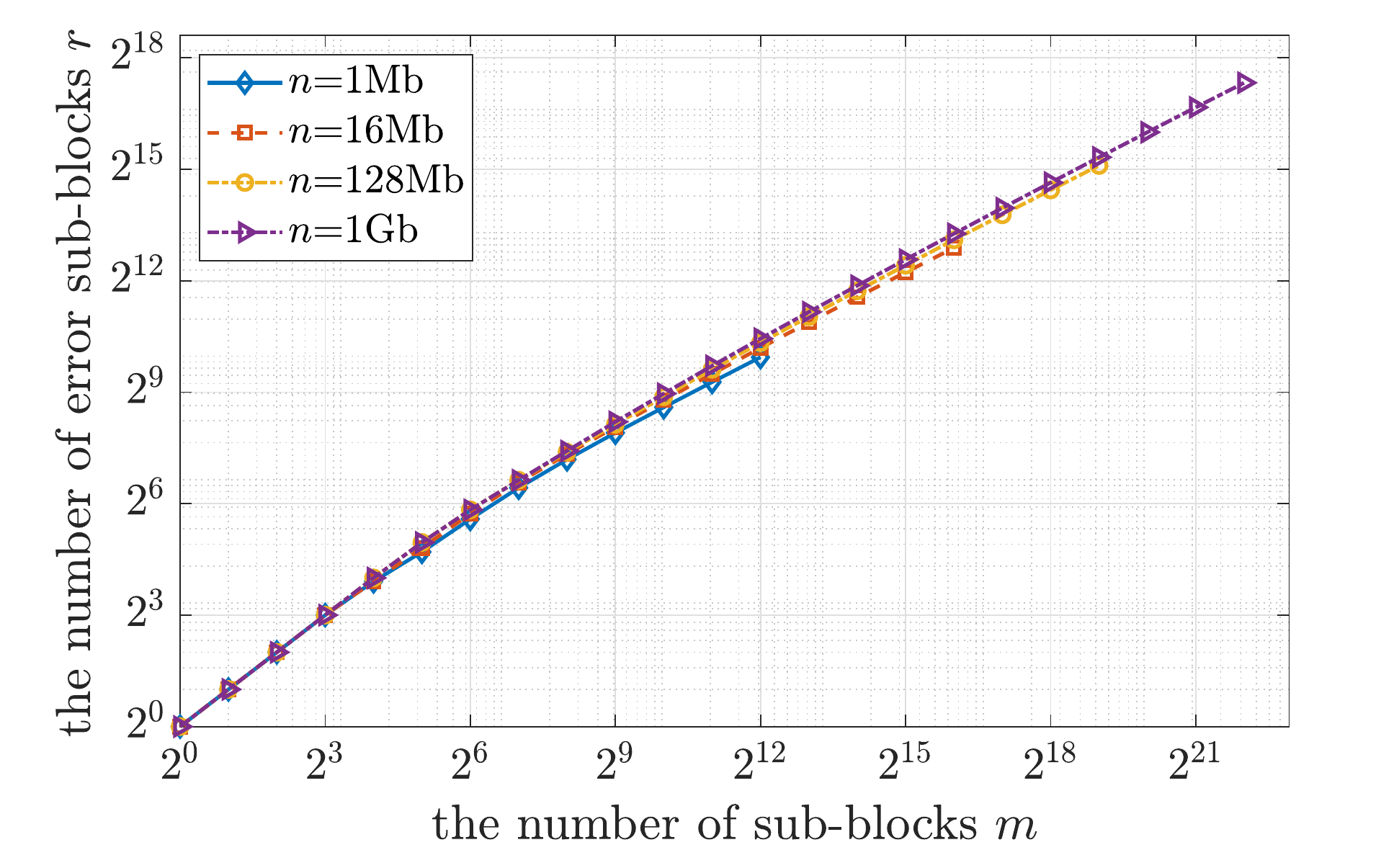}
	\caption{The upper bound of $r$ with different $m$ when QBER is $0.02$ and $md < n H(E_\mu)$\label{fig:error}}
\end{figure}

Assume $P^U_j$ is error probability of the decoded sub-block in the forward reconciliation and the error probability threshold of a sub-block is $\epsilon$, where $j=0, 1, \cdots, m-1$. Thus, the upper bound of $P^U_j$ can be estimated by $P_e(\mathcal{W}_i)$ as 
\begin{equation}
	P^U_j \leq \sum_{i = jn'} ^{\left(j+1\right)n^\prime-1} -v_iP_{e}\left(\mathcal{W}_i\right),
\end{equation}
and the upper bound of the decoded sub-blocks with error bits in forward reconciliation $r$ can be estimated as
\begin{equation}
	r=\left| \left\{ P^U_j \left | \right. j\in \left[0,m \right), P^U_j > \epsilon \right\} \right|.
\end{equation}
With the implementation of upgrading and degrading channel construction of polar codes~\cite{RN106,myConstruction}, the upper bound of $P_e(\mathcal{W}_j)$ is calculated, and the estimated upper bound of $r$ is shown in Fig.\ref{fig:error} with different $m$ when $\epsilon = 10^{-3}$, $E_\mu=0.02$, $d=32$ and $md < n H(E_\mu)$. 

Assume the efficiency of polar codes as $f_\mathrm{I}$, the efficiency of the LDPC codes as $f_\mathrm{II}$. After the acknowledgment reconciliation, the total efficiency of SLA IR scheme is

\begin{equation}
f= \frac{f_\mathrm{I}H_2(E_\mu)+md+m+\varepsilon _\mathrm{f}rn^\prime f_\mathrm{II}H_2\left(E_\mu\right)}{nH_2\left(E_\mu\right)}=f_\mathrm{I}+\frac{m \left(d+1 \right)}{nH_2\left(E_\mu \right)}+\varepsilon _\mathrm{f} f_\mathrm{II} \frac{r}{m},
\end{equation}
where $md$ is the upper bound of leaked information to Eve from the transmitted CRC tag values, extra $m$ bits information may leaked to Eve from the vector $\sigma$ and $\varepsilon _\mathrm{f}rn' f_\mathrm{II}H_2(E_\mu)$ is the syndrome information leaked in the acknowledgment reconciliation.

In the proposed SLA IR scheme, we divide the error correction block to $m$ sub blocks, which will increase the overall efficiency. Without block partition strategy, we have $m=1$ and the efficiency $f_{m=1}$ can be calculated as

\begin{equation}
f_{m=1} = f_\mathrm{I} +\frac{\left(d+1 \right)}{nH_2\left(E_\mu \right)} + \varepsilon _\mathrm{f} f_\mathrm{II}.
\end{equation} 

Thus, given the fixed $f_\mathrm{I}$, the increased efficiency yield $\mathcal{Y}(m)$ of SLA IR scheme with divide the error correction block into $m$ sub blocks can be calculated as

\begin{equation}
	\mathcal{Y}(m)= f_{m=1} -f = -\frac{\left(m-1\right)\left(d+1\right)}{nH\left(E_\mu\right)}+\varepsilon _\mathrm{f} f_\mathrm{II} \frac{m-r}{m}.
\end{equation}

\begin{figure}[htb]
	\centering
	\includegraphics[width=0.8\textwidth]{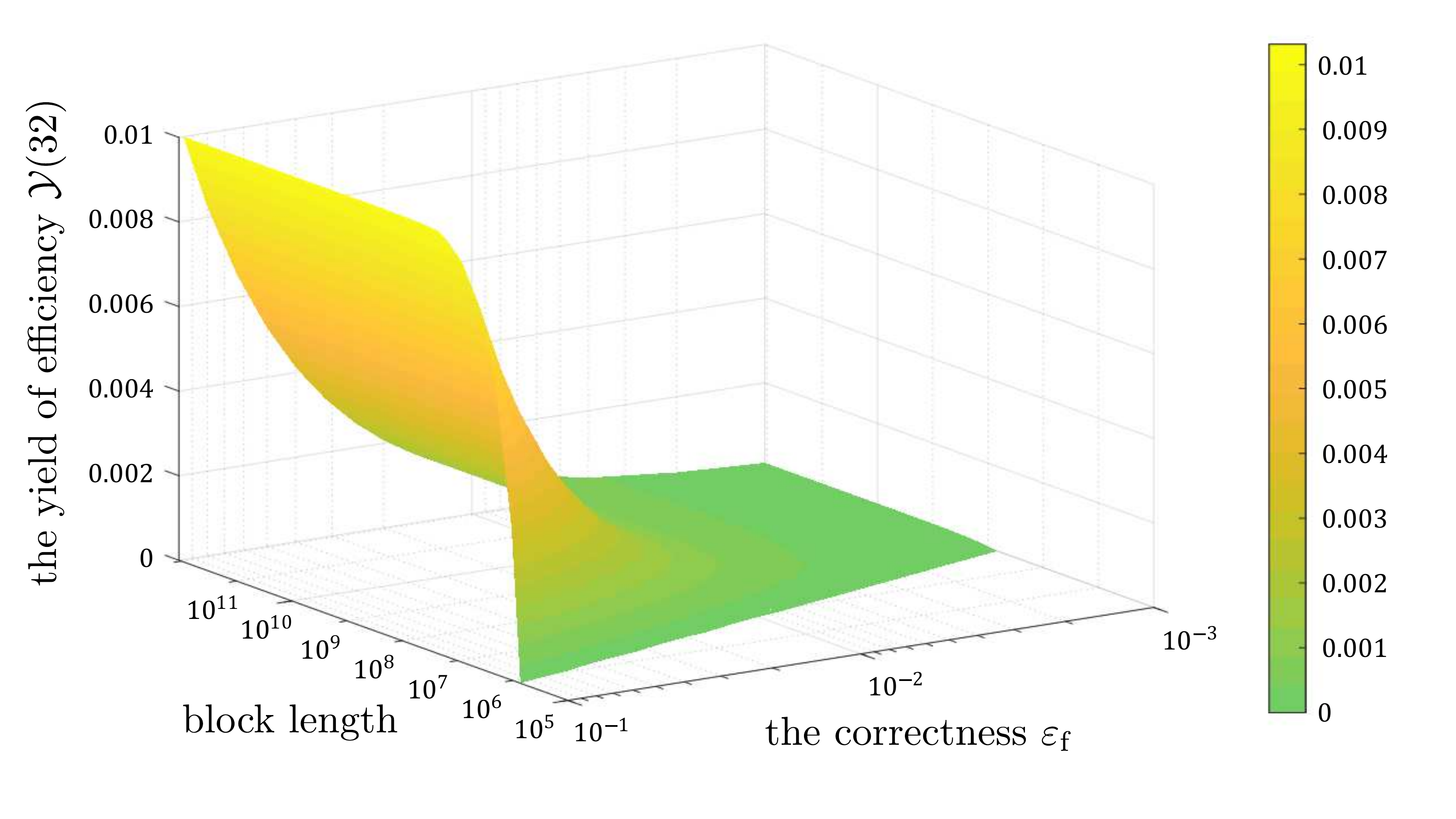}
	\caption{The yield of efficiency $\mathcal{Y}(32)$ when QBER is $0.02$\label{fig:efficiency_yield}}
\end{figure}

The estimation results of $\mathcal{Y}(m)$ are shown in Fig.\ref{fig:efficiency_yield}, where $E_\mu=0.02$, $m=32$. The yield of efficiency increases as block length and $\varepsilon _\mathrm{f}$ increase and approaches to $0.01$ when $\varepsilon_\mathrm{f}$ equals $0.1$ and block length is larger than $10^8$. According to equation (\ref{eq:f}), the yield of efficiency will lead to higher final secure key rates of QKD systems.

\section{Results}
We have implemented the \textbf{S}hannon \textbf{l}imit \textbf{a}pproached (SLA) IR scheme with the block checked (BC) SCL decoder. Afterwards, a series of experiments have been conducted to evaluate the efficiency of the SLA IR scheme with the limitation of $\varepsilon_\mathrm{f} \leq 0.01$. In the experiments, the upgrading and degrading channel construction of polar codes~\cite{RN106,myConstruction} is used to determine the frozen vector of polar codes. The number of sub-blocks $m$ and length of CRC $d$ are both set as $32$ and the list size of BC-SCL decoder $l$ is set as $16$, so that $\varepsilon$-correctness of SLA scheme is calculated as the level of $10^{-8}$ with $\varepsilon _\mathrm{II}=10^{-6}$. Meanwhile, the correction threshold of LDPC~\cite{RN159} is directly used to evaluate the efficiency of the acknowledgment reconciliation.

The SLA IR scheme is tested for $10000$ times each round with QBER ranging from $0.01$ to $0.12$ with step of $0.01$, block length of \SI{1}{Mb}, \SI{16}{Mb}, \SI{128}{Mb}. The experimental results of the $\varepsilon_\mathrm{f}$, the reconciliation efficiency $f$ and average yield $\gamma$ are shown in Table.\ref{tab:result}. Especially, the leaked information used for calculating the efficiency is accumulated from all tests instead of one test for the different error sub-blocks in the forward reconciliation.
 
\begin{table}[!htb]
	\centering
	\caption{The experimental result of the SLA IR scheme\label{tab:result}}
	\begin{tabular}{c|ccc|ccc|ccc}
		\hline
		\multirow{2}*{$E_\mu$}&\multicolumn{3}{c|}{$n$=\SI{1}{Mb}}&\multicolumn{3}{c|}{$n$=\SI{16}{Mb}}&\multicolumn{3}{c}{$n$=\SI{128}{Mb}} \\
		\cline{2-10}
		& $f$ & $\varepsilon_\mathrm{f}$& $\gamma$ & $f$ & $\varepsilon_\mathrm{f}$& $\gamma$ & $f$ & $\varepsilon_\mathrm{f} $& $\gamma$ \\
		\hline
		0.01&1.205 &0.0164 &0.903 &1.114 &0.0032 &0.910 &1.091 &$\leq 10^{-4}$ &0.912 \\
		0.02&1.146 &0.0050 &0.838 &1.085 &0.0138 &0.847 &1.073 &$\leq 10^{-4}$ &0.848\\
		0.03&1.124 &0.0163 &0.782 &1.087 &0.0005 &0.789 &1.062 &0.0011 &0.794 \\
		0.04&1.116 &0.0072 &0.730 &1.072 &0.0048 &0.740 &1.059 &0.0033 &0.743  \\
		0.05&1.107 &0.0046 &0.683 &1.070 &0.0022 &0.694 &1.055 &$\leq 10^{-4}$ &0.698\\
		0.06&1.099 &0.0040 &0.640 &1.062 &0.0050 &0.652 &1.049 &0.0067 &0.657 \\
		0.07&1.101 &0.0012 &0.597 &1.066 &$0.0004$ &0.610 &1.050 &$\leq 10^{-4}$ &0.616  \\
		0.08&1.104 &0.0026 &0.556 &1.064 &$0.0001$ &0.572 &1.048 &$\leq 10^{-4}$ &0.579  \\
		0.09&1.092 &0.0037 &0.523 &1.056 &0.0007 &0.539 &1.044 &0.0015 &0.544 \\
		0.1&1.083 &0.0064 &0.492 &1.062 &$\leq 10^{-4}$ &0.502 &1.042 &$\leq 10^{-4}$ &0.511  \\
		0.11&1.079 &0.0024 &0.461 &1.057 &$\leq 10^{-4}$ &0.472 &1.039 &0.0050 &0.481\\
		0.12&1.072 &0.0043 &0.433 &1.056 &$\leq 10^{-4}$ &0.441 &1.037 &0.0013 &0.451 \\
		\hline
	\end{tabular}
\end{table}

The efficiency $f$ of the SLA IR scheme is $1.205$, $1.114$, $1.091$ when the block length is \SI{1}{Mb}, \SI{16}{Mb}, \SI{128}{Mb} respectively. When the block length increases to \SI{128}{Mb}, the $f$ and $\gamma$ of our SLA IR scheme are much more efficient than the previous polar codes-based IR schemes shown in Table.~\ref{otherwork}. Meanwhile, the efficiency increases and the $\varepsilon _\mathrm{f}$ decreases as block length are increased to \SI{1}{Mb}, \SI{16}{Mb} and \SI{128}{Mb}. Moreover, the SLA IR scheme runs around $167$ hours on a personal computer with the block length $n=$ \SI{1}{Gb}, $E_\mu=0.02$, resulting the efficiency of $1.055$ and $\varepsilon _\mathrm{f}$ less than $10^{-2}$. As we shown, performance of polar codes-based IR schemes can be improved by increasing the block lengths, however,  the implementation of large-scale decoders will result in huge computational complexity, which may destroys the system availability. Therefore, with limited block lengths, our SLA IR scheme can be performed to further improve both the reconciliation efficiency and the correctness of QKD systems.

\section{Conclusion}

In this article, we propose a \textbf{S}hannon-\textbf{l}imit \textbf{a}pproached (SLA) information reconciliation (IR) scheme based on polar codes in quantum key distribution systems, which achieves high reconciliation efficiency and decreases the overall IR failure probability to $10^{-8}$. The proposed SLA IR scheme mainly consists of two phase: the forward reconciliation phase and the acknowledgment reconciliation phase. In the forward reconciliation phase, the sifted key is divided into sub-blocks and performed with the improved block checked successive cancellation list (BC-SCL) decoder, where errors can be efficient located and corrected in each sub-block. Afterwards, the additional acknowledgment reconciliation phase is performed to the failure corrected sub-blocks. The experimental results show that the overall failure probability of SLA IR scheme is decreased to $10^{-8}$ and the efficiency is improved to 1.091 with the IR block length of \SI{128}{Mb}. Therefore, with limited block lengths, our SLA IR scheme can be performed to further improve both the reconciliation efficiency and the correctness of QKD systems. The SLA IR scheme achieves the efficiency of $1.055$ with quantum bit error rate of $0.02$, when the input scale length increased to \SI{1}{Gb}, which is hundred times larger than the state-of-art implemented polar codes-based IR schemes.

\section*{Acknowledgements}

This work was supported in part by the National Natural Science Foundation of China under Grant No. 61972410 and the research plan of National University of Defense Technology under Grant No. ZK19-13.

\section*{Author contributions}
BYT and BL proposed the scheme, performed the experiments, wrote the paper and contributed equally. This work was conceived by BL and WRY, supervised by WRY and co-supervised by CQW. All authors reviewed the manuscript.

\section*{Additional Information}
Competing Interests: The authors declare no competing interests. Correspondence and requests for materials should be addressed to BL or WRY.

\bibliographystyle{unsrt}
\bibliography{ref}

\end{document}